\documentclass[10pt,conference]{IEEEtran}

\ifCLASSINFOpdf
   \usepackage[pdftex]{graphicx}
\else
   \usepackage[dvips]{graphicx}
\fi

\ifCLASSOPTIONcomsoc
  \usepackage[caption=false,font=normalsize,labelfont=sf,textfont=sf]{subfig}
\else
  \usepackage[caption=false,font=footnotesize]{subfig}
\fi

\usepackage{footnote}
\usepackage{nopageno}
\usepackage{float}
\usepackage{url}
\usepackage{amssymb,amsfonts}
\usepackage{amsmath}

\usepackage{graphicx}
\usepackage{mathptmx}  
\usepackage{epstopdf}
\usepackage{textcomp}
\usepackage{xcolor}
\usepackage{xspace}
\usepackage[absolute,showboxes]{textpos}

\usepackage{multirow}
\usepackage{multicol}
\usepackage{array} 
\usepackage{cite}
\usepackage{algorithm,algorithmic}
\usepackage{amsmath}
\usepackage[cmintegrals]{newtxmath}

\usepackage[utf8]{inputenc}
\usepackage{blindtext}
\usepackage{csquotes}

\begin{document}
\pagenumbering{gobble}

\title{DeepXplain: XAI-Guided Autonomous Defense \\ Against Multi-Stage APT Campaigns}

\author{

Trung V. Phan and Thomas Bauschert \\
\IEEEauthorblockA{Chair of Communication Networks, Technische Universit{\"a}t Chemnitz,  09126 Chemnitz, Germany}
Email: trung.phan-van@etit.tu-chemnitz.de, thomas.bauschert@etit.tu-chemnitz.de 

\\
\textcolor{blue}{This paper is currently under review for IEEE GLOBECOM 2026.}
}

% make the title area
\maketitle

\begin{abstract}
Advanced Persistent Threats (APTs) are stealthy, multi-stage attacks that require adaptive and timely defense. While deep reinforcement learning (DRL) enables autonomous cyber defense, its decisions are often opaque and difficult to trust in operational environments. This paper presents \textit{DeepXplain}, an explainable DRL framework for stage-aware APT defense. Building on our prior DeepStage model~\cite{DeepStage}, DeepXplain integrates provenance-based graph learning, temporal stage estimation, and a unified XAI pipeline that provides structural, temporal, and policy-level explanations. Unlike post-hoc methods, explanation signals are incorporated directly into policy optimization through evidence alignment and confidence-aware reward shaping. To the best of our knowledge, DeepXplain is the first framework to integrate explanation signals into reinforcement learning for APT defense. Experiments in a realistic enterprise testbed show improvements in stage-weighted F1-score (0.887 to 0.915) and success rate (84.7\% to 89.6\%), along with higher explanation confidence (0.86), improved fidelity (0.79), and more compact explanations (0.31). These results demonstrate enhanced effectiveness and trustworthiness of autonomous cyber defense.
\end{abstract}

\IEEEpeerreviewmaketitle

\pagestyle{headings}
\setcounter{page}{1}
\pagenumbering{arabic}

\section{Introduction}
\label{sec:introduction}

Advanced Persistent Threats (APTs)~\cite{APTDetectionSurvey1} remain one of the most challenging classes of cyber attacks against enterprise networks. Unlike opportunistic malware, APT campaigns are stealthy, low-and-slow, and multi-stage, evolving through reconnaissance, initial compromise, privilege escalation, lateral movement, command-and-control, and exfiltration. Their distributed and temporally extended nature makes them difficult to detect and contain using traditional rule-based or signature-based defenses, which lack the ability to reason over causal system interactions and long-term attack progression.

Recent advances in deep reinforcement learning (DRL) have enabled the development of autonomous cyber-defense agents capable of adapting to dynamic threats. In our prior work, we proposed \textit{DeepStage}~\cite{DeepStage}, a stage-aware autonomous defense framework that models enterprise activity using fused provenance graphs and learns mitigation policies via DRL. By combining graph-based representation learning and temporal stage estimation, DeepStage improves defense effectiveness against multi-stage APT campaigns. However, similar to many DRL-based systems, it operates as a black box: while it produces effective decisions, it does not provide interpretable explanations for its predictions or actions.

This limitation is particularly critical in security operations. Defense actions such as host isolation, credential revocation, and traffic blocking can incur significant operational disruption and therefore require validation by human analysts. Without interpretable evidence, autonomous defense policies remain difficult to trust, audit, and deploy in practice. Recent advances in explainable reinforcement learning (XRL)~\cite{Milani2024XRL} emphasize the need for transparency in sequential decision-making systems, particularly in safety-critical domains. These studies highlight that DRL agents often lack interpretability, which limits their adoption in real-world settings where reliability and accountability are essential. In parallel, explainable graph neural network (GNN) methods~\cite{GNNExplainer} have emerged as effective tools for identifying influential nodes, edges, and subgraphs that drive model predictions. Such capabilities provide a strong foundation for interpreting provenance-based security analytics. However, existing approaches primarily focus on static prediction tasks and do not establish a direct connection between graph-level explanations and sequential decision-making in reinforcement learning \cite{Milani2024XRL,GNNExplainer}.

In this paper, we propose \textit{DeepXplain}, an XAI-guided extension of DeepStage~\cite{DeepStage} for autonomous defense against multi-stage APT campaigns. DeepXplain introduces a unified explanation pipeline that captures structural evidence from provenance graphs, temporal attribution over attack evolution, and feature-level policy attribution for defense decisions. Unlike conventional post-hoc approaches, these explanation signals are incorporated directly into reinforcement learning through evidence-alignment regularization and confidence-aware reward shaping. This tightly couples explanation with policy learning, enabling the agent to make decisions that are both effective and interpretable. We evaluate DeepXplain in a realistic enterprise testbed using CALDERA-driven APT scenarios~\cite{mitre_caldera}. Experimental results show that DeepXplain improves the average stage-weighted F1-score from 0.887 to 0.915 and increases mitigation success rate from 84.7\% to 89.6\%. In addition, it produces more reliable and concise explanations, achieving higher explanation confidence, compactness and fidelity compared to the post-hoc baseline.

\section{Related Work}
\label{sec:related_work}

\subsection{Explainable Reinforcement Learning}
Recent research on explainable reinforcement learning (XRL) has emphasized the need to make sequential decision-making policies interpretable, particularly in safety- and mission-critical domains. Milani \textit{et al.}~\cite{Milani2024XRL} provide a recent comprehensive survey of XRL and organize the literature around feature attribution, policy summarization, and explanation generation. Their study highlights that most XRL methods remain domain-agnostic and are rarely integrated into the optimization objective itself. Despite this progress, the application of XRL to cyber defense remains limited. Existing autonomous cyber-defense agents typically emphasize mitigation performance, policy convergence, or robustness, but do not explicitly constrain learned policies to align with interpretable evidence. As a result, current RL-based cyber-defense systems often remain difficult to validate and trust in operational settings.

\subsection{Explainable Graph Neural Networks}
With the widespread adoption of graph neural networks (GNNs) for structured data analysis, explainability has become a critical research direction for understanding model decisions. Existing methods aim to identify the most influential nodes, edges, or subgraphs that contribute to a given prediction. For instance, GNNExplainer~\cite{GNNExplainer} learns soft masks over graph structures to maximize the mutual information between a subgraph and the model output, providing instance-level explanations. Recent surveys further systematize this area. Li \textit{et al.}~\cite{Li2025GNNExplainability} categorize GNN explanation methods and highlight key evaluation criteria, including correctness, robustness, usability, and compactness. Similarly, Dai \textit{et al.}~\cite{Dai2024TrustworthyGNN} emphasize that explainability is a fundamental component of trustworthy GNNs, alongside robustness and privacy. Despite these advances, existing approaches are primarily designed for static tasks such as node classification or graph prediction. They do not directly address dynamic and security-critical settings, where graph structures evolve over time and decisions must be taken sequentially. This limitation motivates the need for a unified framework that bridges provenance-graph explanation with policy learning.

\subsection{Autonomous Cyber Defense and RL-Based Mitigation}
Autonomous cyber defense has recently attracted significant attention as defenders seek to respond to machine-speed attacks using learning-based agents. Le \textit{et al.}~\cite{Le2024AutomatedAPTDefense} propose a recent RL-based framework for automated APT defense using attack-graph risk-based situation awareness. Their approach shows that reinforcement learning can improve adaptive mitigation under staged attack conditions, but it relies on abstract attack-graph representations and does not expose interpretable evidence for individual decisions. Similarly, recent work on entity-based reinforcement learning for autonomous cyber defense~\cite{EntityRL2024} explores more realistic defense states, but still focuses primarily on action effectiveness rather than explainability.  These findings motivate the need for defense agents whose decisions are not only effective but also transparent and evidence-driven.

%More broadly, Vyas \textit{et al.}~\cite{Vyas2025ACDSurvey} survey realistic autonomous cyber network defense and conclude that deployment-oriented systems still face substantial challenges in trust, evaluation realism, and operator acceptance.

\subsection{Provenance-Based Security Analytics}
Recent provenance-based intrusion detection systems have shown strong potential for detecting stealthy and multi-stage attacks. FLASH~\cite{Rehman2024FLASH} is a recent representative system that applies graph representation learning to system provenance graphs for scalable intrusion detection. Its results demonstrate the effectiveness of provenance-aware GNN models in capturing causal attack structure beyond flat log features. However, the focus remains on detection rather than autonomous mitigation. More recently, Bilot \textit{et al.}~\cite{Bilot2025PIDSAnalysis} present a comprehensive analysis of provenance-based intrusion detection systems and show that, despite strong reported detection performance, many existing systems face practical challenges related to reproducibility, deployment realism, and analyst usability. This is particularly important for our setting: provenance-based systems can provide rich causal evidence, but without interpretable and action-oriented reasoning they remain difficult to operationalize in autonomous defense workflows.

\subsection{Research Gap}
Taken together, recent work reveals three key gaps. First, XRL methods improve policy transparency but are rarely integrated with cyber-defense agents operating in partially observable, graph-structured environments. Second, explainable GNN research provides strong tools for extracting structural evidence, but does not generally connect that evidence to downstream defense policies. Third, autonomous cyber-defense and provenance-based IDS systems have advanced detection and mitigation capabilities, yet recent analyses show persistent challenges in trust, deployment realism, and analyst interpretability. These limitations motivate the need for a unified framework that links structural explanations with sequential decision making for interpretable APT defense.

\begin{figure*}
    \centering
    \includegraphics[width=0.83\linewidth]{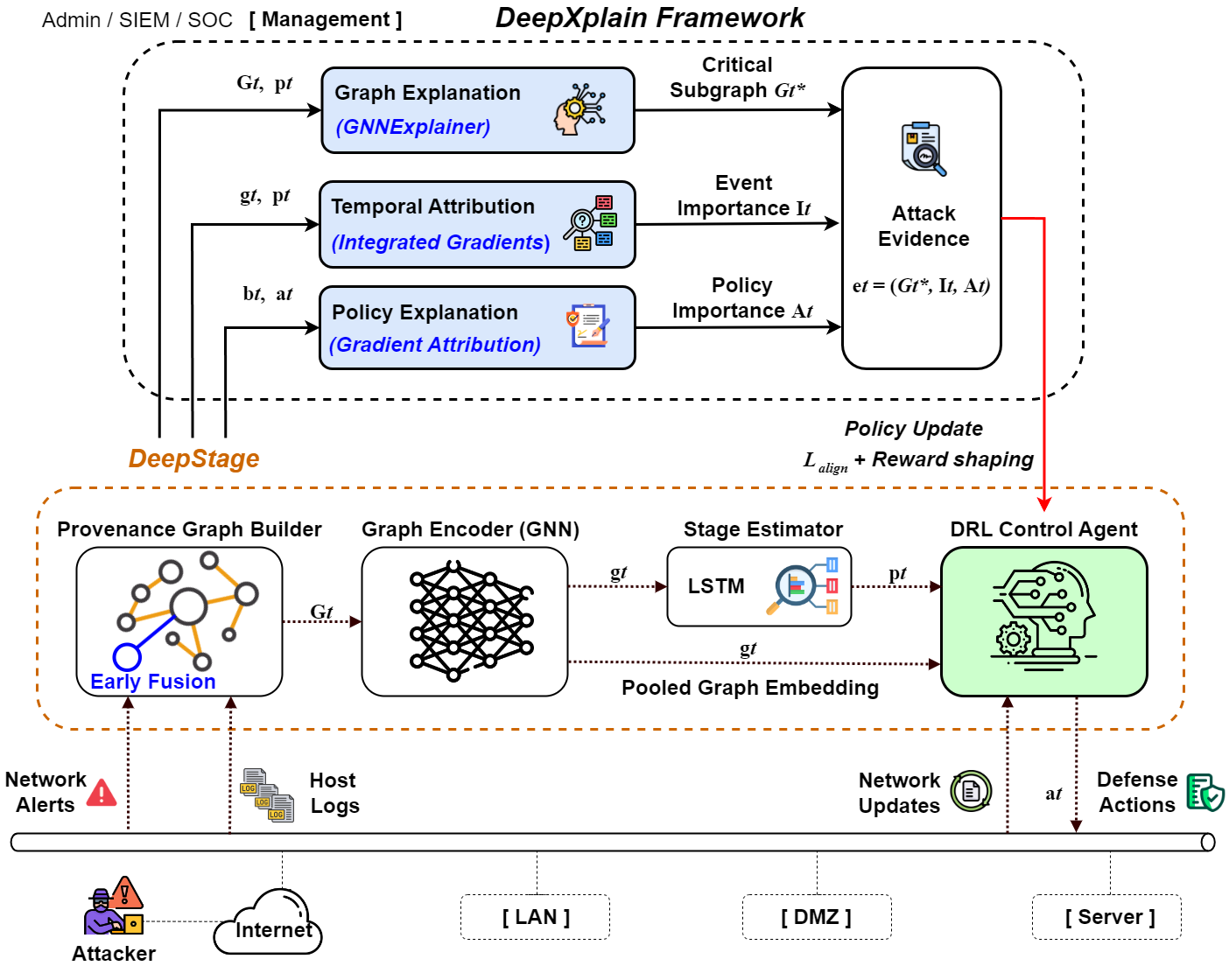}
    \caption{Data and control flow of the DeepXplain framework. The red feedback path highlights how explanation signals are incorporated into policy optimization.}
    \label{fig:DeepXplain_framework_flow}
\end{figure*}

\section{XAI-Guided Autonomous APT Defense}
\label{sec:xai_framework}
This section presents \textit{DeepXplain}, an XAI-guided extension of the DeepStage framework~\cite{DeepStage} for autonomous defense against multi-stage APT campaigns, as illustrated in Figure~\ref{fig:DeepXplain_framework_flow}. While DeepStage enables adaptive defense through deep reinforcement learning, its decision process remains largely opaque, limiting trust and operational deployment.

DeepXplain addresses this limitation by tightly integrating explainable artificial intelligence (XAI) into both inference and decision making. Specifically, it unifies provenance-graph explanation, temporal attribution, and policy-level interpretation, and incorporates these explanation signals directly into policy optimization. This design enables the learned defense policy to be not only effective but also evidence-driven, interpretable, and reliable in security-critical environments.

\subsection{System Model}

Enterprise telemetry from host and network monitoring systems is transformed into a provenance graph $G_t = (V_t, E_t)$, where nodes $V_t$ represent system entities (e.g., processes, files, sockets), and edges $E_t$ encode causal relationships such as process spawning or network communication. This representation captures both structural dependencies and event causality across the system.

A graph neural network (GNN) encodes the provenance graph as $g_t = f_{\mathrm{GNN}}(G_t)$, where $g_t \in \mathbb{R}^{d}$. The embedding $g_t$ summarizes complex interactions within the system into a compact representation suitable for learning.

To capture temporal evolution, the sequence of embeddings is processed by a Stage Estimator: $p_t = f_{\mathrm{LSTM}}(g_1,\dots,g_t)$, where $p_t \in \mathbb{R}^{K}$. The vector $p_t$ represents the probability distribution over $K$ APT stages, enabling temporal reasoning about attack progression.

Since the true system state is partially observable, we adopt a POMDP formulation. The belief state is updated as $b_t = f_{\mathrm{LSTM}}(b_{t-1}, o_t)$ with observation $o_t = [g_t, p_t, a_{t-1}]$, where $b_t$ integrates historical observations and actions to approximate the latent system state. The policy then selects a defense action $a_t \sim \pi_\theta(a|b_t)$, allowing context-aware mitigation strategies.

\subsection{XAI Pipeline}

To enhance interpretability, DeepXplain introduces an XAI module that operates on intermediate representations of the model. At each time step, the module receives
\begin{equation}
\mathcal{X}_t = \{G_t, g_t, p_t, b_t, a_t\},
\end{equation}
which jointly capture structural, temporal, and decision-level information.

The module produces an explanation signal
\begin{equation}
e_t = (G_t^*, I_t, A_t),
\end{equation}
where $G_t^*$ identifies critical graph structures, $I_t$ highlights important time steps, and $A_t$ explains the selected action. This unified representation enables comprehensive reasoning across multiple dimensions of the defense process.

\subsection{Graph-Based Explanation}

To explain attack-stage predictions, we identify a subgraph that preserves the model prediction:
\begin{equation}
G_t^* =
\arg\max_{G' \subseteq G_t}
P(\hat{k}_t | G'), \quad \hat{k}_t = \arg\max_k p_t^{(k)}.
\end{equation}
This formulation seeks the minimal structural evidence sufficient to maintain the predicted stage.

We adopt GNNExplainer~\cite{GNNExplainer}, which learns soft masks
\begin{equation}
M_V \in [0,1]^{|V_t|}, \quad M_E \in [0,1]^{|E_t|},
\end{equation}
assigning importance scores to nodes and edges. These scores highlight critical attack patterns such as suspicious process chains or lateral movement paths.

\subsection{Temporal Attribution}
APT attacks evolve over time, making it essential to identify critical temporal events. We compute temporal attribution as
\begin{equation}
I_i =
\left|
\frac{\partial P(\hat{k}_t)}{\partial g_i}
\right|_1.
\end{equation}
This measures how strongly each past embedding influences the current prediction. The normalized importance $\tilde{I}_i = \frac{I_i}{\sum_j I_j}$ forms a probability distribution over time steps, enabling identification of key transitions in the attack lifecycle.

\subsection{Explainable Defense Policy}
To interpret defense decisions, we analyze the sensitivity of the policy to the belief state:
\begin{equation}
A_i =
\left|
\frac{\partial \pi(a_t|b_t)}{\partial b_t^{(i)}}
\right|.
\end{equation}
This reveals which features—such as high attack-stage probability or anomalous activity—drive the selected action. The normalized attribution is defined as $\phi_{\mathrm{policy}} = \frac{A_t}{\|A_t\|_1}$, which provides a comparable importance distribution.

To align structural explanations with policy reasoning, we project graph explanations into belief space:
\begin{equation}
g_t^* = \sum_{v \in V_t} w(v) h_v,
\end{equation}
where $w(v)$ is node importance and $h_v$ is the node embedding. We then compute $\phi_{\mathrm{XAI}} =
\frac{W_g g_t^* + W_p p_t}{\|W_g g_t^* + W_p p_t\|_1}$, which integrates structural evidence and stage information into the same feature space as the policy, enabling direct comparison.

\subsection{XAI-Guided Policy Optimization}

We incorporate explanation signals into policy learning through an alignment loss:
\begin{equation}
\mathcal{L}_{align} =
\|\phi_{\mathrm{policy}} - \phi_{\mathrm{XAI}}\|_2^2,
\end{equation}
which encourages the policy to focus on features supported by interpretable evidence.

To quantify explanation reliability, we define
\begin{equation}\label{Confidence_Function}
\mathrm{Conf}(e_t)
=
\alpha C_{\mathrm{graph}}
+
\beta C_{\mathrm{temp}}
+
\gamma C_{\mathrm{policy}},
\end{equation}
where each term captures a different aspect of explanation quality. Here, $C_{\mathrm{graph}} =
\frac{\sum_{v \in V_t^*} w(v)}{\sum_{v \in V_t} w(v)}$ measures structural completeness, $C_{\mathrm{temp}} =
1 - \frac{-\sum_i \tilde{I}_i \log \tilde{I}_i}{\log t}$ measures temporal concentration, and $C_{\mathrm{policy}} =
\frac{\phi_{\mathrm{policy}}^\top \phi_{\mathrm{XAI}}}
{\|\phi_{\mathrm{policy}}\|_2 \|\phi_{\mathrm{XAI}}\|_2}$ measures consistency between explanation and policy.

Finally, the augmented objective is
\begin{equation}\label{DRL_ObjectiveFunction}
J(\theta)
=
J_{RL}(\theta)
-
\lambda_1 \mathcal{L}_{align}
+
\lambda_2 \mathrm{Conf}(e_t),
\end{equation}
where $\lambda_1$ enforces evidence alignment and $\lambda_2$ rewards reliable explanations. In practice, this objective is optimized using PPO, where the augmented reward replaces the standard reward during policy updates. This formulation tightly integrates explanation and decision making, enabling DeepXplain to learn policies that are not only effective but also interpretable and robust against complex multi-stage APT behaviors.

\section{Performance Evaluation}
\label{sec:evaluation}

This section evaluates the proposed \textit{DeepXplain} framework with respect to both \emph{defense effectiveness} and \emph{explanation quality}. Since DeepXplain is built on top of DeepStage, we inherit the same enterprise testbed, telemetry pipeline, attack scenarios, and defense action space used in our prior work~\cite{DeepStage}. Unless otherwise stated, all components unrelated to explainability follow the same configuration as DeepStage to isolate the impact of the proposed XAI-guided extensions.

\begin{table*}[t]
\centering
\caption{Per-stage F1-score and overall defense effectiveness comparison.}
\label{tab:defense_results}
\small
\begin{tabular}{lccccccccc}
\hline
Method & Recon & InitAcc & PrivEsc & LatMov & C2 & Exfil & Avg-F1 & Success Rate (\%) \\
\hline
Risk-Aware DRL    & 0.75 & 0.73 & 0.70 & 0.71 & 0.76 & 0.72 & 0.728 & 68.5 \\
DeepStage         & 0.91 & 0.88 & 0.85 & 0.87 & 0.92 & 0.89 & 0.887 & 84.7 \\
DeepXplain        & \textbf{0.93} & \textbf{0.91} & \textbf{0.89} & \textbf{0.90} & \textbf{0.94} & \textbf{0.92} & \textbf{0.915} & \textbf{89.6} \\
\hline
\end{tabular}
\end{table*}

\subsection{Experimental Setup}

\subsubsection{Enterprise Testbed and Data Sources}
Following DeepStage~\cite{DeepStage}, experiments are conducted in a realistic enterprise testbed logically segmented into four zones: local area network (LAN), demilitarized zone (DMZ), server zone, and management zone. Host-level telemetry is collected from endpoint monitoring tools (e.g., Auditd), while network-level events are captured by Zeek-based sensors. These events are normalized and fused into provenance graphs, where nodes represent system entities and edges encode causal interactions.

To emulate realistic multi-stage APT campaigns, we use CALDERA-driven adversarial playbooks~\cite{mitre_caldera} covering reconnaissance, initial compromise, privilege escalation, lateral movement, command-and-control, and exfiltration. Benign background activities are executed concurrently to create realistic noise and increase detection difficulty. The same attack-stage definitions, telemetry collection interval, and graph construction window used in DeepStage are retained here to ensure a fair comparison. All reported results are averaged over 10 independent runs.

\subsubsection{Model Configuration}
DeepXplain reuses the core DeepStage backbone, including the GNN encoder, LSTM-based Stage Estimator, and PPO-based defense policy. The graph encoder produces an embedding $g_t \in \mathbb{R}^{d}$, and the Stage Estimator outputs $p_t \in \mathbb{R}^{K}$. The belief state $b_t \in \mathbb{R}^{d_b}$ is updated recurrently using the same observation structure as DeepStage.

For the XAI components, the graph explanation module is implemented using GNNExplainer~\cite{GNNExplainer}. The explainer learns node and edge masks for each graph instance over 100 optimization steps with learning rate $10^{-2}$. We retain the top-$m$ explanation nodes after masking, where $m$ is selected to preserve at least 90\% of the cumulative node importance in $G_t^*$. Temporal attribution is computed using gradient-based sensitivity over the sequence of graph embeddings, and policy attribution is computed from the policy network gradients with respect to the belief-state features.

\subsubsection{XAI-Guided Optimization Parameters}
The XAI-guided objective introduced in Section~\ref{sec:xai_framework} includes an alignment regularization term and a confidence-based reward term in Equation~\eqref{DRL_ObjectiveFunction}. In our experiments, the alignment coefficient is set to $\lambda_1 = 0.1$, which provides a moderate constraint encouraging policy attention to align with explanation-derived evidence while preserving policy flexibility. The explanation confidence weight is set to $\lambda_2 = 0.05$, allowing confidence-aware reward shaping without overwhelming the original defense objective. These values are chosen from the ranges discussed in Section~\ref{sec:xai_framework}, namely $\lambda_1 \in [0.01,0.5]$ and $\lambda_2 \in [0.01,0.3]$, and were found to provide stable training. For the explanation confidence score in Equation~\eqref{Confidence_Function}, we set $(\alpha,\beta,\gamma)=(0.4,0.2,0.4)$. This weighting prioritizes structural completeness and policy-evidence consistency, while still accounting for temporal concentration.

\subsubsection{Baselines}
We compare DeepXplain against the following baselines:
\begin{itemize}
    \item \textit{Risk-Aware DRL}~\cite{AutomatedAPTDefense}: A reinforcement learning-based defense framework that leverages attack-graph risk modeling to minimize cumulative security risk.
    \item \textit{DeepStage}~\cite{DeepStage}: The original stage-aware autonomous defense framework without XAI guidance.
    \item \textit{DeepStage + Post-hoc XAI}: DeepStage augmented with explanation modules for interpretability only, without incorporating explanation signals into policy learning.
\end{itemize}
%The comparison with \textit{DeepStage + Post-hoc XAI} isolates the gain from XAI-guided optimization beyond merely generating explanations.

\subsection{Results Analysis}
\subsubsection{Defense Effectiveness}

Table~\ref{tab:defense_results} summarizes both the per-stage and overall defense effectiveness of all methods under the same evaluation setting as DeepStage. DeepXplain consistently achieves the best performance across all APT stages, attaining the highest average stage-weighted F1-score of 0.915, compared to 0.887 for DeepStage and 0.728 for Risk-Aware DRL. The improvement is uniform across all stages and is particularly significant during privilege escalation and lateral movement, where accurate situational awareness and timely intervention are critical to preventing attack propagation. These results highlight the benefit of incorporating explanation-guided signals, which enable the agent to focus on causally relevant system behaviors rather than spurious correlations. Notably, these gains are obtained despite the already strong performance of DeepStage, indicating that explanation-aware learning contributes beyond representation learning alone.

The comparison with Risk-Aware DRL further demonstrates the limitations of static, risk-oriented state representations. While risk-aware approaches capture global attack surfaces, they rely on coarse attack-graph abstractions and lack the fine-grained, temporal, and causal context necessary to model multi-stage APT behavior. As a result, their performance degrades in later stages of the attack, where precise reasoning over system dynamics is essential.

Compared with DeepStage, DeepXplain achieves a notable improvement in overall mitigation effectiveness, increasing the success rate from 84.7\% to 89.6\%. The mitigation success rate is defined as the percentage of attack episodes in which the defense agent prevents the APT from reaching critical stages such as command-and-control or exfiltration. These gains indicate that integrating XAI into the learning process not only enhances interpretability but also leads to more reliable and efficient defense decisions. By aligning policy learning with interpretable evidence, DeepXplain enables the agent to prioritize compact, temporally consistent, and causally meaningful signals, thereby improving both robustness and response quality in complex APT scenarios.

\subsubsection{Defense Responsiveness}

\begin{figure}
\centering
\includegraphics[width=1.0\linewidth]{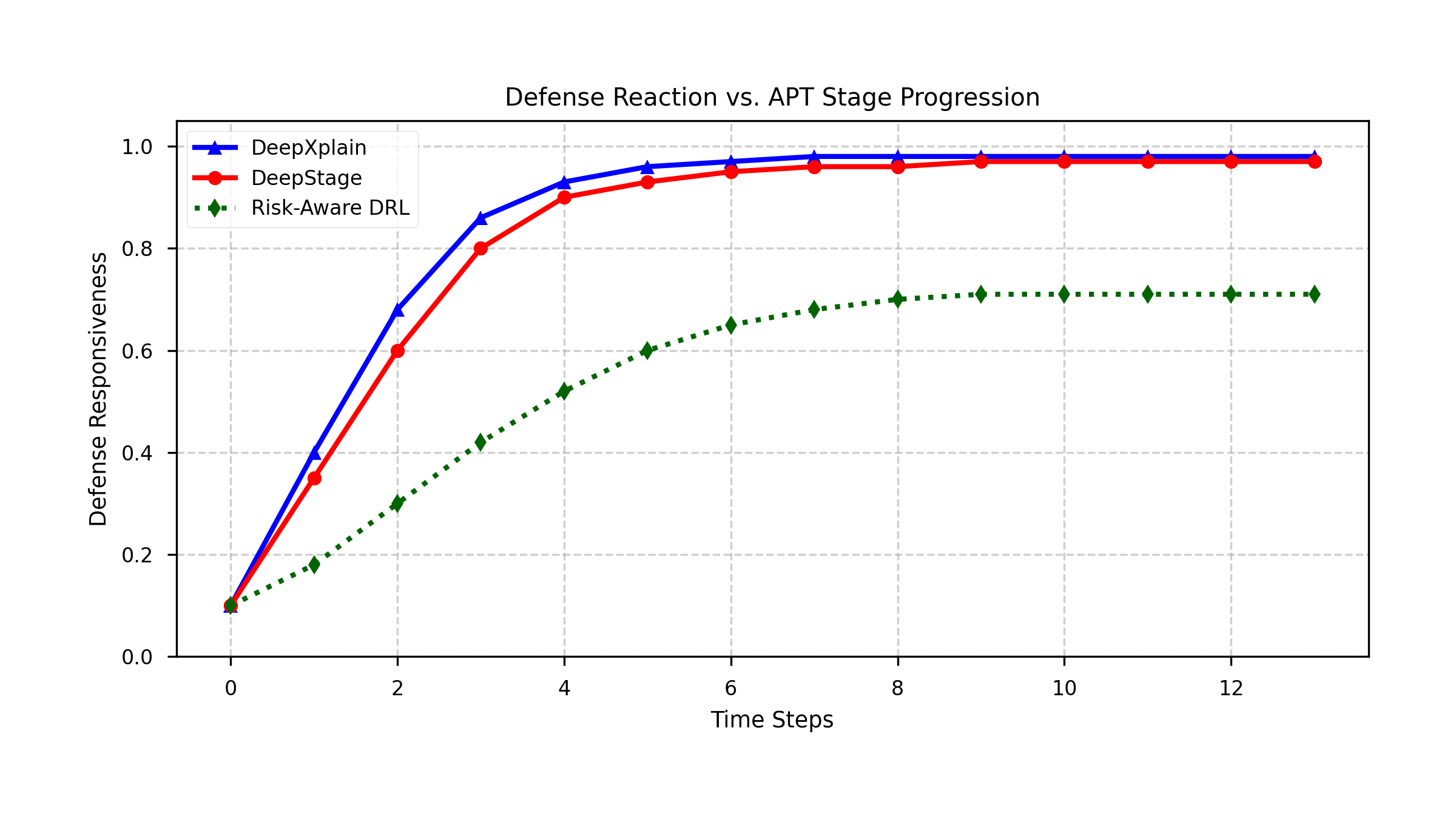}
\caption{Defense reaction versus APT stage progression.}
\label{fig:stage_timeline}
\end{figure}

Figure~\ref{fig:stage_timeline} illustrates the defense responsiveness across the APT lifecycle. DeepXplain consistently reacts faster and maintains higher responsiveness than all baselines. At early attack transitions ($t=3$), DeepXplain achieves a responsiveness of 0.86, outperforming DeepStage (0.80, +7.5\%) and Risk-Aware DRL (0.42, +104.8\%). During critical escalation ($t=5$), it reaches 0.96, compared to 0.93 for DeepStage (+3.2\%) and 0.60 for Risk-Aware DRL (+60.0\%). At convergence, DeepXplain stabilizes at approximately 0.98, exceeding DeepStage (0.97) and significantly outperforming Risk-Aware DRL (0.71, +38.0\%). The advantage is most evident during the transition from privilege escalation to lateral movement, where rapid containment is crucial. This improvement is driven by explanation-guided policy optimization, which enables earlier detection of stage transitions and more timely, evidence-consistent responses.

\begin{figure}
\centering
\includegraphics[width=1.0\linewidth]{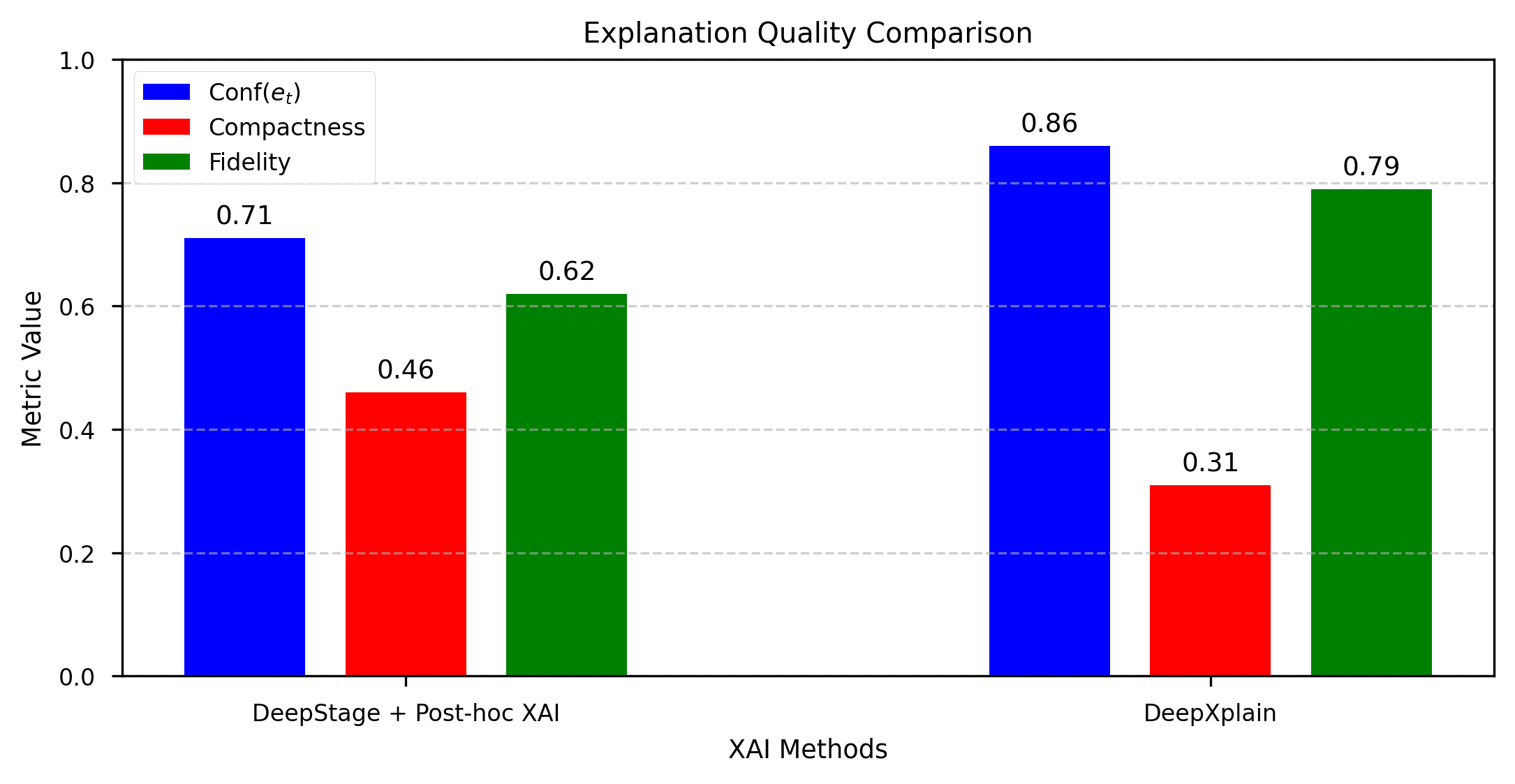}
\caption{Explanation quality comparison.}
\label{fig:explanation_quality}
\end{figure}

\subsubsection{Explanation Quality}

Figure~\ref{fig:explanation_quality} evaluates the quality of the generated explanations from two complementary perspectives: confidence, compactness and fildelity. DeepXplain consistently outperforms DeepStage + Post-hoc XAI across all metrics. Specifically, DeepXplain achieves a higher explanation confidence score (0.86 vs. 0.71, +21.1\%), indicating that the extracted evidence is more consistent and reliable. Compactness measures the conciseness of explanations, defined as the proportion of important graph elements required to explain a decision. Lower compactness values indicate that the explanation focuses on a smaller, more relevant subset of nodes and edges. DeepXplain produces more compact explanations (0.31 vs. 0.46, $-32.6\%$), demonstrating its ability to localize critical attack patterns while filtering out noisy dependencies. In addition, we evaluate fidelity, which measures how well the generated explanations reflect the true decision-making behavior of the model. Fidelity is assessed by examining how much the model’s prediction confidence degrades when the identified explanatory components are removed. DeepXplain achieves a higher fidelity score (0.79 vs. 0.62, +27.4\%), indicating that its explanations capture the actual causal factors driving the model’s predictions, rather than superficial correlations.

These results highlight a fundamental limitation of post-hoc explanation methods. While they provide interpretability after training, they do not influence the learning process and therefore cannot prevent reliance on spurious correlations. In contrast, DeepXplain integrates explanation signals directly into policy optimization, enforcing consistency between policy attention and graph-derived evidence. This results in explanations that are more concise, reliable, and causally grounded, leading to decisions that are both interpretable and operationally meaningful.

%These results highlight a fundamental limitation of post-hoc explanation methods. While they provide interpretability after training, they do not influence the learning process and therefore cannot prevent reliance on spurious correlations. In contrast, DeepXplain integrates explanation signals directly into policy optimization, enforcing consistency between policy attention and graph-derived evidence. This results in explanations that are more concise, reliable, and causally grounded, leading to decisions that are both interpretable and operationally meaningful. In other words, explanation in DeepXplain is not merely descriptive, but actively improves policy learning.

\begin{table}[t]
\centering
\caption{Ablation study on XAI-guided optimization.}
\label{tab:ablation_results}
\begin{tabular}{lcc}
\hline
Variant & Avg. Stage-F1 & Conf$(e_t)$ \\
\hline
DeepXplain w/o $\mathcal{L}_{align}$ & 0.900 & 0.79 \\
DeepXplain w/o $\mathrm{Conf}(e_t)$ & 0.910 & 0.74 \\
DeepXplain (full) & \textbf{0.915} & \textbf{0.86} \\
\hline
\end{tabular}
\end{table}

\subsubsection{Ablation Study}

The ablation results in Table~\ref{tab:ablation_results} quantify the contribution of each component in the proposed XAI-guided objective. Removing the alignment loss $\mathcal{L}_{align}$ reduces the average stage-F1 from 0.915 to 0.900 ($-1.6\%$) and decreases explanation confidence from 0.86 to 0.79 ($-8.1\%$). This indicates that aligning policy attribution with graph-based evidence is essential for ensuring that the agent focuses on causally relevant features, thereby improving both robustness and interpretability. On the other hand, removing the confidence-based reward $\mathrm{Conf}(e_t)$ results in a smaller drop in stage-F1 (0.915 to 0.910, $-0.5\%$) but a more significant reduction in explanation confidence (0.86 to 0.74, $-14.0\%$). This suggests that the confidence term primarily enhances the reliability and consistency of explanations by encouraging the agent to favor actions supported by concentrated and high-quality evidence across different attack scenarios.

Overall, the full DeepXplain model achieves the best trade-off between defense effectiveness and explanation quality. The ablation study demonstrates that the alignment loss improves decision accuracy by enforcing evidence consistency, while the confidence reward enhances explanation reliability. Their combination enables DeepXplain to jointly optimize performance and interpretability, rather than treating explanation as a purely post-hoc artifact.

\section{Conclusion}
This paper presented \textit{DeepXplain}, an XAI-guided extension of DeepStage for autonomous defense against multi-stage APT campaigns. The framework integrates provenance-based graph reasoning, temporal stage inference, and deep reinforcement learning with a unified explanation pipeline. Unlike post-hoc approaches, DeepXplain incorporates explanation signals directly into policy optimization through evidence alignment and confidence-aware reward shaping. Experimental results demonstrate improvements in both defense effectiveness and interpretability, achieving higher stage-wise performance and more reliable explanations. Future work will explore human-in-the-loop security operations and the integration of large language models for enhanced decision support.

%\section{Conclusion}
%This paper presented \textit{DeepXplain}, an XAI-guided extension of DeepStage for autonomous defense against multi-stage APT campaigns. The framework integrates provenance-based graph reasoning, temporal stage inference, and deep reinforcement learning with a unified explanation pipeline combining graph, temporal, and policy attribution. Unlike post-hoc approaches, DeepXplain incorporates explanation signals directly into policy optimization through evidence alignment and confidence-aware reward shaping. Experimental results demonstrate that DeepXplain improves both defense effectiveness and interpretability, achieving higher stage-wise performance and more compact and reliable explanations than existing methods. More broadly, DeepXplain represents a shift from post-hoc explainability to explanation-aware learning for autonomous cyber defense. Future work will explore human-in-the-loop security operations and the integration of large language models for enhanced interpretability and decision support.

%\section*{Acknowledgment}
%This work has been performed in the framework of the SUSTAINET-Advance project, funded by the German BMFTR (ID:16KIS2280).

\bibliographystyle{ieeetr}
\bibliography{References.bib}

\end{document}